\documentclass[longtitle,preprint,5p,times,twocolumn]{elsarticle}

\usepackage{amssymb}

\usepackage{caption}
\usepackage{xspace}
\usepackage{longtable}
\usepackage{pdflscape}
\usepackage{multirow}
\usepackage{ragged2e}
\usepackage{booktabs}
\usepackage{hyperref}

\usepackage[mathlines,modulo]{lineno}
\usepackage[utf8]{inputenc}
\usepackage{mdwlist}
\usepackage[T1]{fontenc}
\usepackage{textcomp}
\usepackage{tgpagella}
\usepackage{amsmath}
\usepackage{hyperref}
\usepackage{lineno}
\usepackage{lipsum}
\usepackage{sidecap}

\usepackage{graphicx}
\usepackage{axodraw4j}
\usepackage{bm}

{\begin{list}{}%
	{\setlength{\leftmargin}{#1}}%
	\item[]%
}
{\end{list}}
{\begin{list}{}%
	{\setlength{\leftmargin}{-0.5#1}}%
	\item[]%
}
{\end{list}}

\begin{document}

\begin{frontmatter}

\title{Search for $e\to \tau$ Charged Lepton Flavor Violation at the EIC with the ECCE Detector}

\def\theaffn{\arabic{affn}} 
%
%
%
%
%
%
\author[SDU]{J.-L.~Zhang}
\author[UNGeorgia]{S.~Mantry}

\author[MoreheadState]{J.~K.~Adkins}
\author[RIKEN,RBRC]{Y.~Akiba}
\author[UKansas]{A.~Albataineh}
\author[ODU]{M.~Amaryan}
\author[Oslo]{I.~C.~Arsene}
\author[MSU]{C. Ayerbe Gayoso}
\author[Sungkyunkwan]{J.~Bae}
\author[UVA]{X.~Bai}
% ...

%

% ...
\affiliation[MoreheadState]{organization={Morehead State University},
	 city={Morehead},
	 state={KY},
	 }
 
% ...
\affiliation[MSU]{organization={Mississippi State University},
	 city={Mississippi State},
	 state={MS},
	 country={USA}} 
 
% ...
\affiliation[ODU]{organization={Old Dominion University},
	 city={Norfolk},
	 state={VA},
	 country={USA}} 
 
% ...
\affiliation[RIKEN]{organization={RIKEN Nishina Center},
	 city={Wako},
	 state={Saitama},
	 country={Japan}} 
 
% ...
\affiliation[RBRC]{organization={RIKEN BNL Research Center},
	 city={Upton},
	 state={NY},
	 country={USA}} 
	 
\affiliation[SDU]{organizaton={Shandong University},
     city={Qingdao},
     state={Shandong},
     country={China}}
     
% ...
\affiliation[Sungkyunkwan]{organization={Sungkyunkwan University},
	 city={Suwon},
	 country={Republic of Korea}} 
 
% ...
\affiliation[UNGeorgia]{organization={University of North Georgia},
     cite={Dahlonega}, 
     state={GA},
     country={USA}}
     
% ...
\affiliation[UKansas]{organization={Unviersity of Kansas},
	 city={Lawrence},
	 state={KS},
	 country={USA}} 
 
% ...
\affiliation[Oslo]{organization={University of Oslo},
	 city={Oslo},
	 country={Norway}} 
 
% ...
\affiliation[UVA]{organization={University of Virginia},
	 city={Charlottesville},
	 state={VA},
	 country={USA}} 
 
% ...

\begin{abstract}

The recently approved Electron-Ion Collider (EIC) will provide a unique new opportunity for searches of charged lepton flavor violation (CLFV) and other new physics scenarios. In contrast to the $e \leftrightarrow \mu$ CLFV transition for which very stringent limits exist, there is still a relatively large discovery space for the $e \to \tau$ CLFV transition, potentially to be explored by the EIC. With the latest detector design of ECCE (EIC Comprehensive Chromodynamics Experiment) and projected integral luminosity of the EIC, 
we find the $\tau$-leptons created in the DIS process $ep\to \tau X$ are expected to be identified with high efficiency. A first ECCE simulation study, restricted to the 3-prong $\tau$-decay mode and with limited statistics for the Standard Model backgrounds, estimates that the EIC will be able to improve the current exclusion limit on $e\to \tau$ CLFV by an order of magnitude.
\end{abstract}

\end{frontmatter}

\linenumbers
\modulolinenumbers[1]

\section{Charged-Lepton Flavor Violation and Leptoquarks}
The discovery of neutrino oscillations provided  conclusive evidence of lepton flavor violation. Lepton flavor violation in the neutrino sector also results in charged lepton flavor violation (CLFV) through loop-suppressed processes such as $\mu \to e \gamma$. However, the resulting predicted CLFV rates are highly suppressed due to the small neutrino masses --  Br$(\mu \to e  \gamma) < 10^{-54}$ -- and are far beyond the reach of any current or planned experiments. On the other hand, many Beyond Standard Models (BSM) scenarios predict CLFV rates  that are both much larger and within reach of ongoing or near-future experiments. For example, supersymmetry-based models predict rates as high as Br$(\mu \to e  \gamma) \sim 10^{-15}$~\cite{Barbieri:1995tw}, while the current experimental limit on the $\mu \to e \gamma$ process already reached Br$(\mu \to e  \gamma) < 4.2\times  10^{-13}$~\cite{MEG:2016leq}. On the other hand, while there have been extensive searches for CLFV processes between the first and second lepton generations, denoted as CLFV(1,2) for brevity, the constraints on CLFV(1,3) processes that involve $e\leftrightarrow \tau$ transitions are weaker by several orders of magnitude. These constraints on CLFV(1,3)~\cite{Gonderinger:2010yn,Cirigliano:2021img} were obtained through searches for $e +  p\to \tau +X$, $\tau\to e\gamma$, and $p+p\to e+\tau + X$ at HERA~\cite{ZEUS:2005nsy,H1:2007dum}, BaBar~\cite{BaBar:2009hkt}, and the LHC~\cite{ATLAS:2018mrn} respectively.
However, there are many BSM scenarios such as grand unified theories with leptoquarks and R-parity violating supersymmetry that predict CLFV(1,3) rates that are enhanced compared to CLFV(1,2) processes, motivating continued searches dedicated for $e\leftrightarrow \tau$ transitions.

\begin{figure}[!h]
\centering
\includegraphics[width=0.4\textwidth]{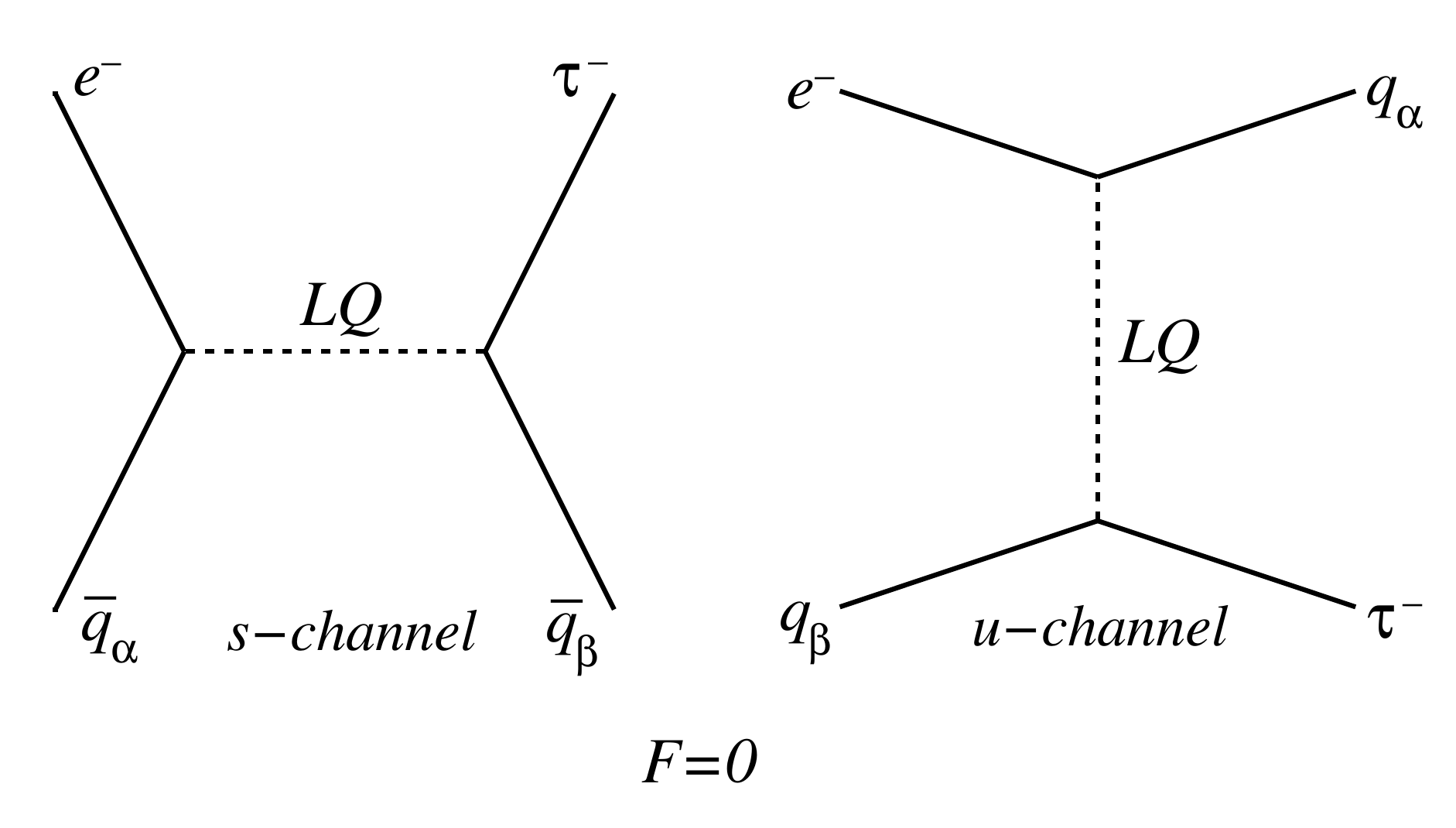}
\includegraphics[width=0.4\textwidth]{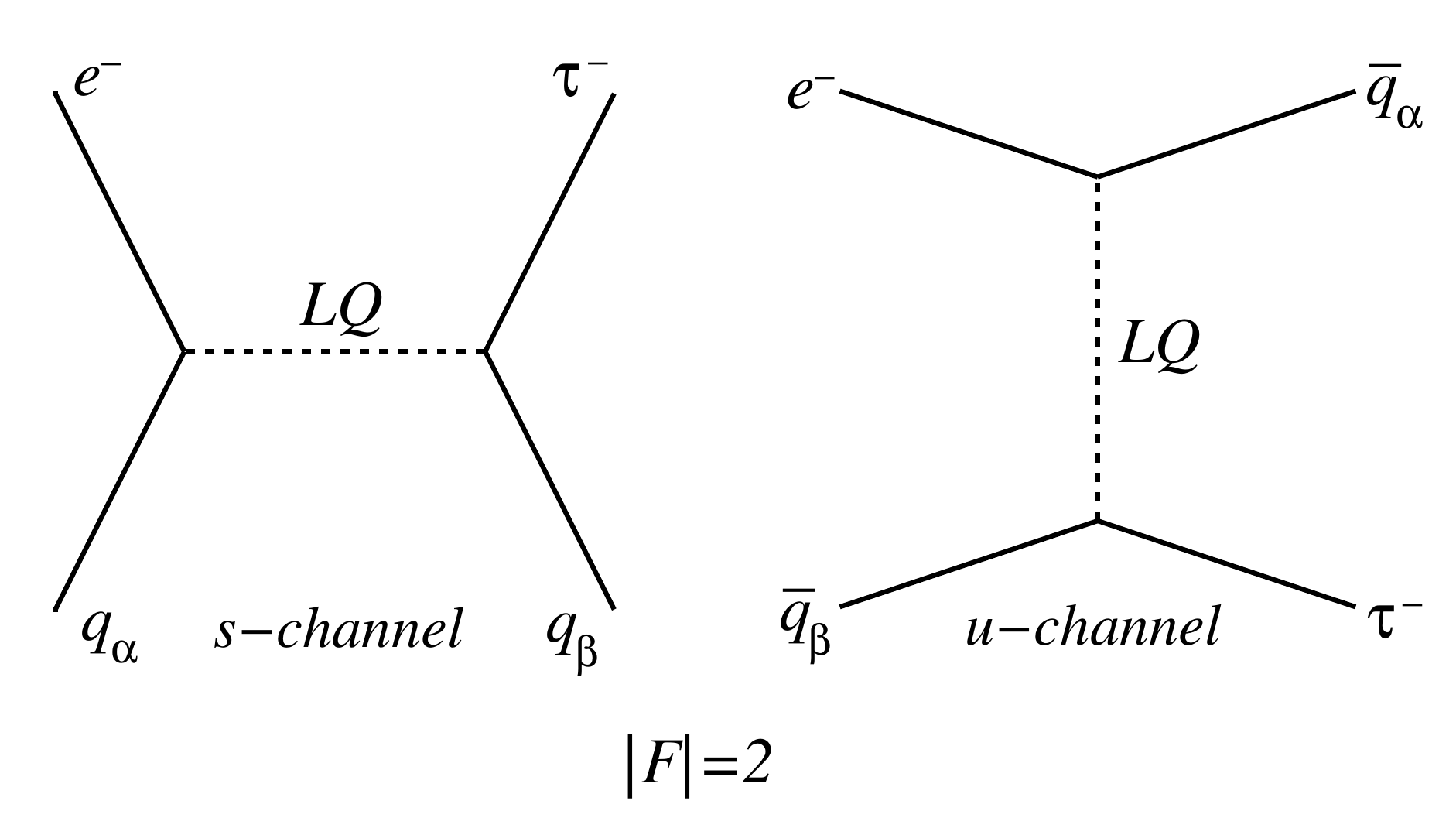}
\caption{From~\cite{Gonderinger:2010yn}: Representative Feynman diagrams for $e\to \tau$ scattering processes via one-leptoquark mediator. The fermionic number $F$ is assumed to be conserved, as in the BRW effective model~\cite{Buchmuller:1986zs}. The partonic cross section is convoluted with the PDF of the initial state (anti)quark of each diagram, and depends on the parameter $\lambda_{1\alpha}\lambda_{3\beta}/M_{LQ}^2$.}\label{fig:etaufeyndiagrams}
\end{figure}

We carry out the simulation analysis based on the design of the ECCE Detector (recommended as Detector 1 by the EIC Detector Proposal Advisory Panel~\cite{DPAPreport2022}), for determining the sensitivity to the CLFV(1,3) process $e+p\to \tau+X$ in the leptoquark framework~\cite{Gonderinger:2010yn,Cirigliano:2021img}, though such analysis could also be performed in the SMEFT framework~\cite{Cirigliano:2021img}. Leptoquarks are color triplet bosons that carry both lepton ($L$) and baryon
($B$) numbers, coupling leptons to quarks and mediating the $e +  p\to \tau +X$ CLFV(1,3) process at tree-level, as shown in Fig.~\ref{fig:etaufeyndiagrams}. The leptoquarks are classified into 14 types~\cite{Buchmuller:1986zs} based on their fermion number $F=3B + L$ ($F = 0$ or $\vert F\vert = 2$), spin (scalar or vector), chiral couplings to leptons (left-handed or right-randed), $SU(2)_L$ representation (singlet, doublet, or triplet), and $U(1)_Y$ hypercharge.  

In the region 
where the leptoquark mass $M_{\rm LQ}$ is much larger than the characteristic energy scale of the experiment, represented by the center-of-mass energy $\sqrt{s}$ and the four-momentum transfer $\sqrt{Q^2}$, the CLFV(1,3) process $e +  p\to \tau +X$ is mediated by a contact interaction and the tree-level cross section for $F=0$ or $|F|=2$ leptoquarks takes the form:
\begin{eqnarray}
\label{eq:elecF0LQcrosssection}
\hspace*{-0.8cm}&&\sigma_{F=0}=\sum_{\alpha,\beta} \frac{s}{32\pi} {\left[  \frac{\lambda_{1\alpha}\lambda_{3\beta}}{M_{LQ}^2}  \right]}^2 \times\nonumber\\
 &&\hspace*{0.8cm}\int dx \int dy \> \big \{ x \bar{q}_\alpha(x,x s) f(y) + x q_\beta(x,-u) g(y)\big \}, \nonumber\\
\hspace*{-0.8cm}&&\sigma_{|F|=2}=\sum_{\alpha,\beta} \frac{s}{32\pi} {\left[ \frac{\lambda_{1\alpha}\lambda_{3\beta}}{M_{LQ}^2} \right] }^2 \times\\
&&\hspace*{0.8cm}\int dx \int dy  \>\big \{ x q_\alpha(x,x s) f(y) + \bar{q}_\beta(x,-u) g(y) \big \},\nonumber
\end{eqnarray}
where $u=x (y-1) s$ with $x$ the Bjorken scaling variable and $y$ the fractional energy loss of the electron in the proton-rest frame. The kinematic $y$-dependent functions $f(y),g(y)$ are $f(y)=1/2, g(y)=(1-y)^2/2$ and 
$f(y)=2(1-y)^2, g(y)=2$ for scalar and vector leptoquarks, respectively. The quantity $\lambda_{1\alpha}\lambda_{3\beta}/(M_{LQ}^2)$ characterizes the strength of the contact interaction. The $\lambda_{ij}$ parameters, assumed to be real numbers for this analysis, denote the leptoquark couplings between the $i$-th lepton generation and $j$-th quark generation.

The ZEUS and H1 experiments at HERA placed upper limits on  $\lambda_{1\alpha}\lambda_{3\beta}/M_{LQ}^2$~\cite{ZEUS:2002dnh,ZEUS:2005nsy,H1:1999dil,H1:2007dum}. The HERA data set corresponded to $\sqrt{s}=300$ and 318~GeV and a total integrated luminosity of up to 130~pb$^{-1}$.  With several orders of magnitude increase in the luminosity, the EIC has the potential to improve upon the HERA limits for both diagonal ($\alpha = \beta$) and off-diagonal ($\alpha\neq \beta$) components, and provide complementary information~\cite{Cirigliano:2021img} to the constraints from BaBar and the LHC.


\section{EIC/ECCE Simulation and Vertex Identification}

The EIC Comprehensive Chromodynamics Experiment (ECCE) detector concept~\cite{ecce-detector-proposal} addresses the full EIC science mission as described in the EIC community White Paper~\cite{Accardi:2012qut} and the 2018 National Academies of Science (NAS) Report~\cite{NAP25171}. It is simultaneously fully capable, low-risk, and cost-effective. ECCE strategically repurposes select components of existing experimental equipment to maximize its overall capabilities within the envelope of planned resources. For example, the central barrel of the detector incorporates the storied 1.4 T BaBar superconducting solenoid, and the sPHENIX barrel hadronic calorimeter currently under construction.

The goal of this work is to study the potential for discovering the $e^-p\to\tau^-X$ CLFV process at the EIC based on realistic detector simulations, and identification of such events over Standard Model (SM) backgrounds that include neutral-current (NC) deep inelastic scattering (DIS), charged-current (CC) DIS, and photoproductions. 
CLFV events leading to a final-state $\tau$ and are characterized by a high-momentum isolated $\tau$ which is balanced by a jet in the transverse plane. Since the $\tau$ will decay into stable particles after a short flying distance of $\sim\mu$m, only its decay products are visible in the detector. 
A critical requirement of CLFV searches is thus the secondary vertex reconstruction performance which relies on the tracking and especially the vertex detectors. The ECCE conceptual design uses state-of-the-art technologies consisting Monolithic Active Pixel Sensor (MAPS) based silicon vertex/tracking subsystem to achieve high precision primary and decay vertex determination.

The three SM background processes affect the CLFV searches as follows: First, the SM NC DIS events are very similar to leptoquark events where the $\tau$ decays to only one charged particle plus neutrinos in the final state, making this channel very difficult to study. 
Secondly, due to the presence of at least one neutrino in all $\tau$-decay channels, significant missing $p_T$ is expected. This feature is similar to the SM CC DIS events. 
The third main background of concern is from photoproduction events, mostly due to their very high yield. In this study, we focus the identification of CLFV candidate events that contain a high-p$_T$ quark-initiated jet along with an isolated and high-p$_T$ $\tau$ which replaces the scattered electrons in the typical NC DIS events, and the rejection of events from all three SM background processes.

Based on the number of the charged particles in the final state, the dominant $\tau$ decay modes can be categorized into ``1-prong'' (one charged particle) and ``3-prong'' (three charged particles). The ``3-prong'' decay modes have a branching ratio of $\sim$15\% while the ``1-prong'' modes have a branching ratio of $\sim$85\%. Although the ``1-prong'' modes have larger branching ratio, they are more demanding to identify.  For example, the ``1-prong'' mode could be one of the two leptonic decays, $\tau\to e \bar{\nu}_e\nu_\tau$ or $\tau\to \mu \bar{\nu}_\mu\nu_\tau$.  If the charged particle is an electron, it is very similar to DIS NC events. If the charged particle is a muon, it will require good muon identification.  Another possible ``1-prong'' mode is $\tau\to \nu_\tau \pi^-$, containing one charged pion, and can be studied in the near future.
In this study, we focus only on searching for the $\tau$ ``3-prong'' decay events. 

The features of ``3-prong'' leptoquark events include: 1) no scattered electron is detected; 2) a high transverse-momentum ($P_{T}$) $\tau$-jet consists of three charged particles within a relatively small cone; 3) all three charged particles originate from a common secondary vertex; 4) a high $P_{T}$ hadronic jet is found back-to-back from the $\tau$-jet; and 5) a $P_{T}$-imbalance caused by the escaped neutrinos which should be part of the candidate $\tau$-jet. In order to simulate such candidate events and the capability to identify them, the leptoquark quark generator LQGENEP~\cite{Bellagamba:2001fk} (version 1.0) is used to produce the signal Monte-Carlo (MC) events, while Djangoh and Pythia generators are used to produce the background DIS and photoproduction MC events, respectively. 
Considering the large mass of leptoquarks, we focus on the highest energy, 18$\times$275 GeV $ep$ collision. For the LQGENEP simulation, the $Q^2$ range is set to $>10$ GeV$^2$ and a default value of leptoquark mass 1.9~TeV from the generator is used, see~\ref{sec:app_LQ_inputs} for input files. Scanning through a series of leptoquark mass values, no visible difference in the characteristics of these signal events was observed at EIC kinematics. The input files for background event generation are given in~\ref{sec:app_LQ_bg_Djangoh} for DIS NC and CC, and~\ref{sec:app_LQ_bg_pythia} for photoproduction.  After passing the generated events through the ECCE GEANT4 
simulation, an analysis algorithm with preliminary selection requirements based on the features of the signal and background events are applied on both leptoquark and SM MC event samples. 

\begin{figure}[!h]
    \centering
    \includegraphics[width=0.48\textwidth]{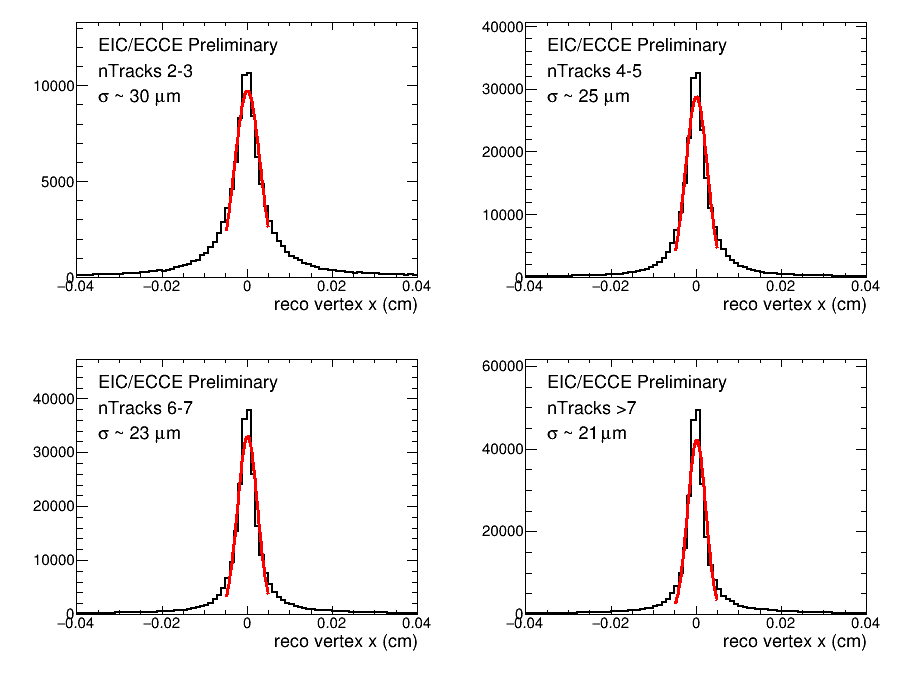}
    \caption{Primary vertex resolution: difference between truth and reconstructed information for different track multiplicity. Only the $x$-component is shown here, while $y$- and $z$-components show similar characteristics but are not shown. }\label{fig:LQ_prVreso}
\end{figure}

As mentioned earlier, a precise identification capability of the interaction vertex is essential for the secondary vertex reconstruction and $\tau$ identification. Figure~\ref{fig:LQ_prVreso} shows the vertex resolution for different track multiplicities, where we see that the ECCE configuration can provide a vertex resolution of $20-30 \mu$m while the decay length of $\tau$ lepton is $\sim$87$\mu$m. Therefore the ECCE vertex resolution is sufficient for identifying $\tau$ decays.

\begin{figure}[!ht]
    \centering
    \includegraphics[width=0.45\textwidth]{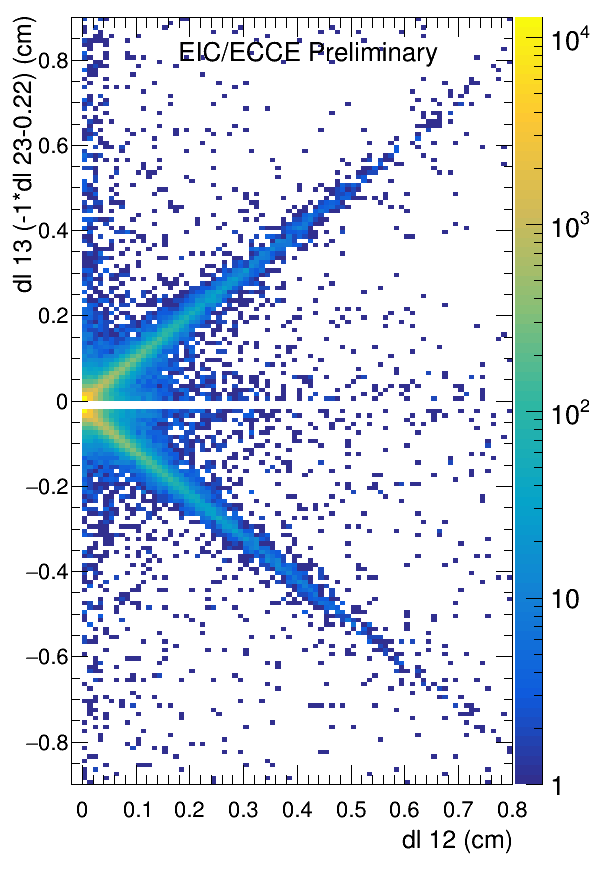}
    \caption{Coincidence among three ``intermediate'' vertices for 3-$\pi$ event identification. The $x$-axis, the positive $y$ and the negative $y$-axis (displaced and direction reversed for clarity), represent the ``intermediate'' vertices from the three pair-combinations $12$, $13$ and $23$, respectively. }
    \label{fig:LQ_vtxCoro}
\end{figure}

To reconstruct the secondary vertex, we first look for 3-$\pi$ candidate events. In the following, the charged pion's tracking information is from the simulated tracking and the vertex detector responses, though particle identification (PID) is based on generator information (namely, perfect PID is assumed). In our algorithm, one track is matched to a second track and the middle point at the closest approach is the candidate secondary vertex position which will be further justified based on the topological structure. For a given 3-$\pi$ candidate event, there are three such pair-combinations and we can reconstruct three ``intermediate'' vertices and the candidate vertex will be the average of all three. Figure~\ref{fig:LQ_vtxCoro} shows the correlation of three intermediate vertices from the three pair combinations where the three decay lengths $dl_{12}$, $dl_{13}$, and $dl_{23}$, as extracted by the distance from the primary to the secondary ``intermediate'' vertices, are shown as the $x$-axis, the positive half of the $y$-axis, and the negative half of the $y$-axis, respectively. There are clearly two bands centralized at lines $y = \pm x$ and the 3-$\pi$ vertex can be identified by requiring either or both correlations. In fact, when combined with vector alignment cuts, coincidence between two of the three ``intermediate'' vertices (either upper or lower half of Fig.~\ref{fig:LQ_vtxCoro}) is usually enough to indicate a ``3-prong" secondary vertex. 

\section{Event Selection}\label{sec:lq_selection}
\begin{figure*}[!h]
    \centering
    \includegraphics[width=0.9\textwidth]{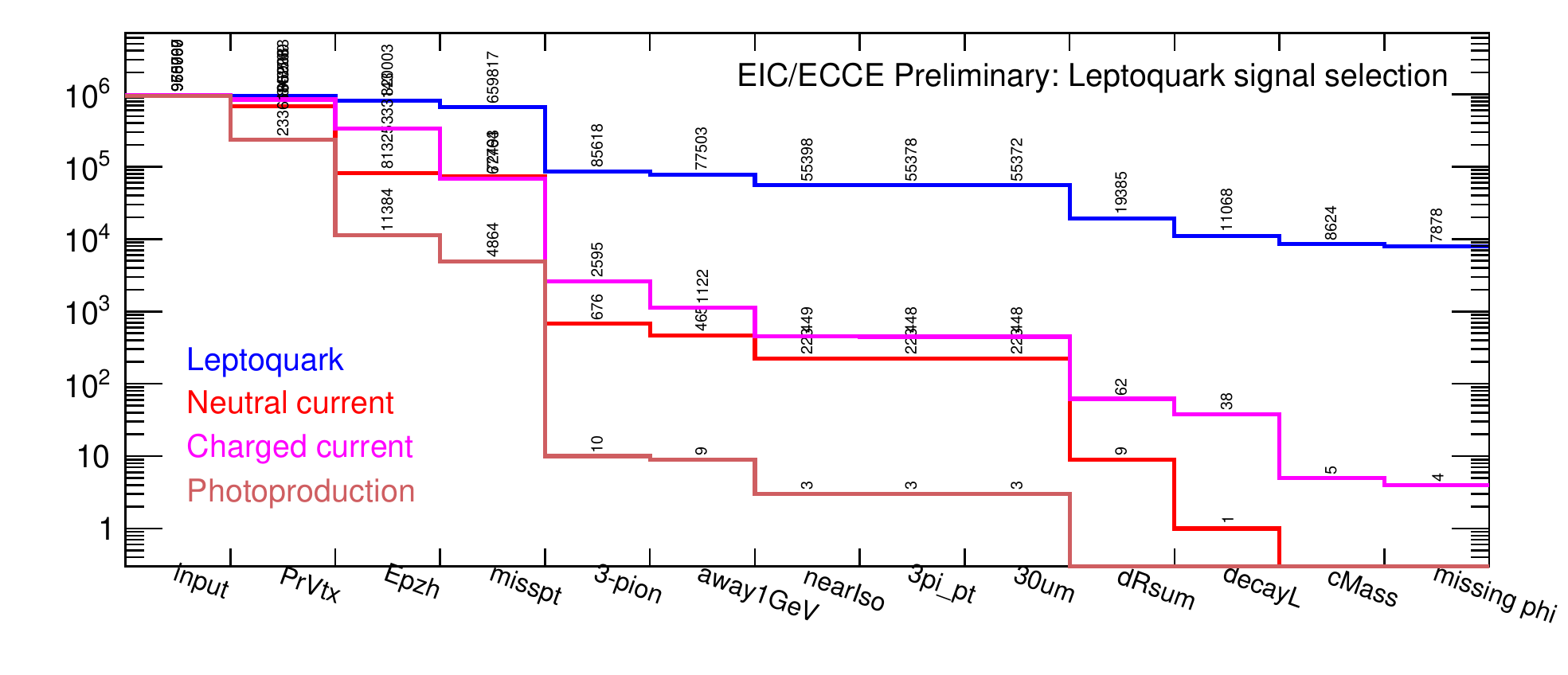} 
    \vspace*{-0.5cm}
    \caption{MC statistics of leptoquark (blue), DIS CC (red), DIS NC (magenta), and photoproduction (orange) events, as ten selection criteria are progressively applied on 1~M input events for each channel. Please see text for details.}
    \label{fig:LQ_cutstat}
\end{figure*}

We used ten selection criteria to identify $e\to\tau$ events and to reject SM backgrounds. Their effects are shown in Fig.~\ref{fig:LQ_cutstat}, where the vertical axis shows how many sample events pass each of the selection criteria, and the horizontal axis are the progressive selections defined as follows: 
\begin{itemize}
\setlength\itemsep{-0.2em}
    \item input: initial input events. We used $10^6$ (1M) MC events for each of leptoquark, DIS NC, DIS CC, and photoproduction processes;
    \item PrVtx: there must be a primary vertex reconstructed; 
    \item Epzh: $\sum_{h}(E-p_{z}) >$ 18 GeV, where $E$ and $p_z$ are the energy and the $z-$component (along the beams) of the 3-momentum of the final state particles, respectively, and the summation is over all detected hadrons;
    \item misspt: 1 $<$ missing $p_T <$ 9 GeV, here the lower limit is to suppress events with small missing $p_T$, e.g. photo-production events, and the upper limit is to suppress DIS events with large missing $p_T$ caused by neutrinos (CC) or miss-detected electrons (NC); 
    \item 3-pion: candidate 3 charged pions are found in a $\Delta R < 1.0$ cone, where $R$ is cone radius in the azimuth($\phi$)-pseudorapidity($\eta$) space, $\Delta R \equiv \sqrt{\Delta\phi^2+\Delta\eta^2}$;
    \item away1GeV: $p_T$ sum of all tracks on the away-side of the candidate 3-$\pi$,  $\sum_{\Delta\phi(-\vec{p}_{3\pi}) < 1.0}p_T$, is $>1$ GeV;
    \item nearIso: 
    $p_T$ sum in a cone around the candidate 3-$\pi$, 
    $\sum_{\Delta R(\vec{p}_{3\pi}) < 1.0}p_T$, is $< 3.0$ GeV;
    \item 3pi\_pt: $p_T$ sum of the 3 charged-pion candidate, $p_{T(3\pi)}$, is $> 3.0$ GeV; 
    \item 30$\mu$m: candidate decay length reconstructed from any pair of the 3 charged pions is $> 30\mu$m; 
    \item dRsum: sum of the ``distances'' (in $\phi-\eta$ space) of the 3 charged pions decay vectors, 
    $\Delta$R$_{1,2}$ + $\Delta$R$_{1,3}$ + $\Delta$R$_{2,3}$, is $< 0.4$. Here the decay vector is defined as starting from the primary vertex and pointing to the secondary vertex;
    \item decayL: average of the reconstructed decay lengths from three pair combinations of the 3-$\pi$ candidate, ($dl_{12}$ + $dl_{13}$ + $dl_{23}$)/3, is $> 0.5$ mm;
    \item cMass: $\sqrt{M_{3\pi}^2+p^2_{3\pi}\mbox{sin}^2\theta}+p_{3\pi}\mbox{sin}^2\theta < 1.8$ GeV, where $\theta$ is the angle between the reconstructed decay direction and the $3\pi$ momentum direction, and $M_{3\pi}$ is the mass reconstructed from the 3-$\pi$~\cite{LHCb:2015}; 
    \item missing phi: missing $p_T$ is azimuthally on the near side of the candidate 3-$\pi$, that is, $\Delta\phi$ between $\vec{p_{3\pi}}$ and $\vec{p_T^{miss}}$ is $< 1.0$. 
\end{itemize}

\begin{figure}[!h]
    \includegraphics[width=0.48\textwidth]{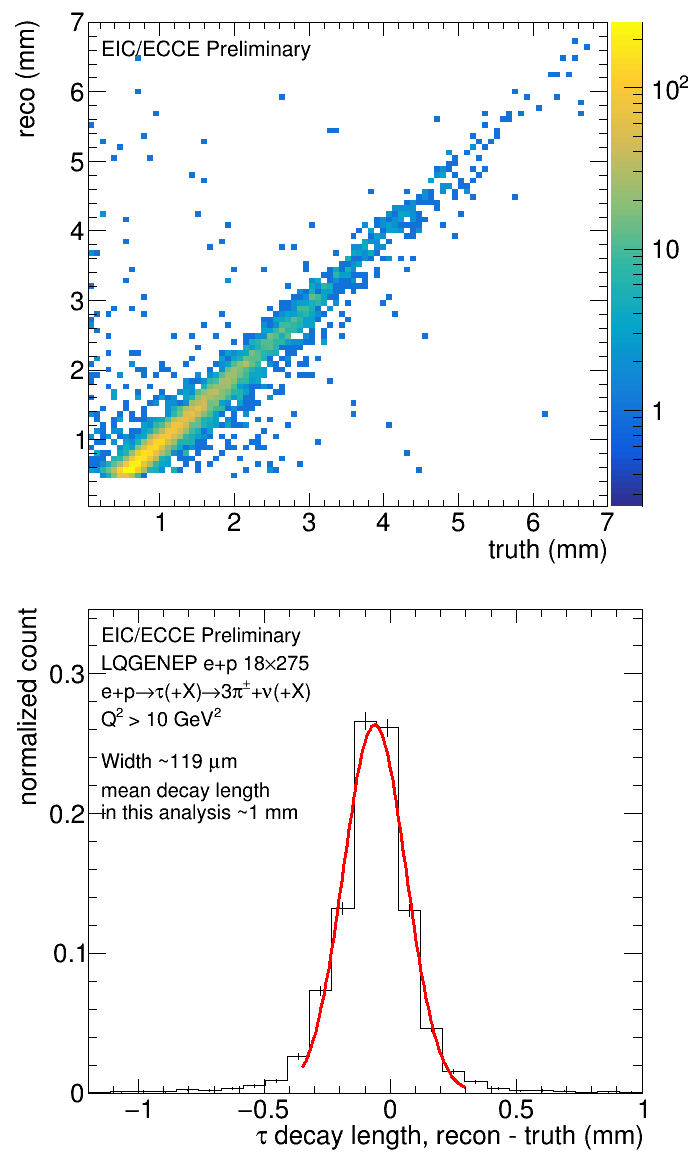}
\hspace*{-0.5cm}
\caption{Top: Reconstructed decay length from Geant4 detector simulation vs. true decay length from generator level. Bottom: difference between reconstructed and true decay length with a Gaussian fit.}
\label{fig:LQ_decaylength}
\end{figure}

From Fig.~\ref{fig:LQ_cutstat}, it can be seen that the $e\to\tau$ events can be effectively selected with this set of preliminary cuts. 
In addition, selections using the decay length are the most discriminating feature of the $\tau$-jet. We illustrate this feature, characterized by the precision of ECCE's vertex detector, in Fig.~\ref{fig:LQ_decaylength}. The left panel shows a comparison between the true decay length from the generator and the decay length reconstructed from tracks at the detector level, while the right panel shows a 119~$\mu$m resolution of the decay length under ECCE configuration, capable for the $\tau$ vertex identification. 

\section{Sensitivity to Leptoquarks}
We can now deduce the sensitivity to the leptoquark signal cross section based on simulations of the 3-prong decay mode (15\% branching ratio) of the $\tau$ lepton, discussed in the last section, and considering  different possible values for the detection efficiency of the other $\tau$ decay modes. In Fig.~\ref{fig:LQ_cutstat}, 1M MC event samples are generated for each of the four processes: the leptoquark mediated signal process $e+p\to\tau +X$, and three background processes, NC DIS, CC DIS, and photoproduction.  After all selection cuts are applied, including the requirement of detecting three pions corresponding to the 3-prong tau decay mode, about 8K (7878) of the original 1M leptoquark signal events remain. For an integrated luminosity of 100~fb~$^{-1}$, the 1M signal events generated corresponds to a signal cross section of 10$^4$fb. Thus, if we assume that only the 3-prong mode has a non-zero detection efficiency, the minimum required cross section for detecting a single signal event in the 3-prong mode is $10^4~{\rm fb}/7878 = 1.3$~fb. On the other hand, if we allow for the other tau decay modes and assume that they can be detected with the same efficiency as the 3-prong decay mode, then the number of candidate events that  now survive selection cuts will be $7878 /(15\%) =52520$. Thus, in this case the minimum required cross section for detecting a single signal event will be $10^4{\rm fb}/52520 = 0.19$~fb. We also consider an intermediate scenario in which, in addition to the 3-prong mode, we allow for the muon and single charge pion decay modes ($\sim$~40~\% branching ratio) and assume they can be detected at half the detection efficiency of the 3-prong decay. In this case the total number of candidate events that now survive all  selection cuts will be $7878 +7878/(15\%)\times (40\%) \times 1/2= 18382$. In this case, the minimum required cross section to detect a single signal event is $10^4{\rm fb}/18382 = 0.54$~fb.


For searches of rare events like leptoquark mediated $e\to\tau$ transitions, the background simulation is more difficult than the signal.
Among the three types of background events, the CC DIS background is easier to estimate because the total cross section is only $\approx 2.3\times 10^4$~fb, and the 1~M CC DIS MC events generated correspond to about 43\% of the expected statistics for 100~fb$^{-1}$ of integrated luminosity. There are 4 CC DIS events that pass the event selection, which can be scaled up to $4/43\%=9$ events for 100~fb$^{-1}$. With more careful optimization, this number can be suppressed even further. On the other hand, the NC DIS and photoproduction backgrounds are much harder to evaluate because their cross sections are much larger and of order $10^7$~fb. Thus, the 1M generated MC events correspond to only $\sim 10^6/(10^7 {\rm fb}\times 100{\rm fb}^{-1} ) = 0.1\%$ of the expected data sample for  100~fb$^{-1}$ of integrated luminosity. With this limited MC event sample, zero NC and photoproduction event satisfies all selection criteria. Thus, at this stage, it is difficult to estimate how many background events will survive the selection if the MC sample is increased by factor $10^3$, which is needed to simulate 100~fb$^{-1}$ of NC DIS and photoproduction data. 

\begin{figure}[!h]
    \centering
    \includegraphics[width=0.48\textwidth]{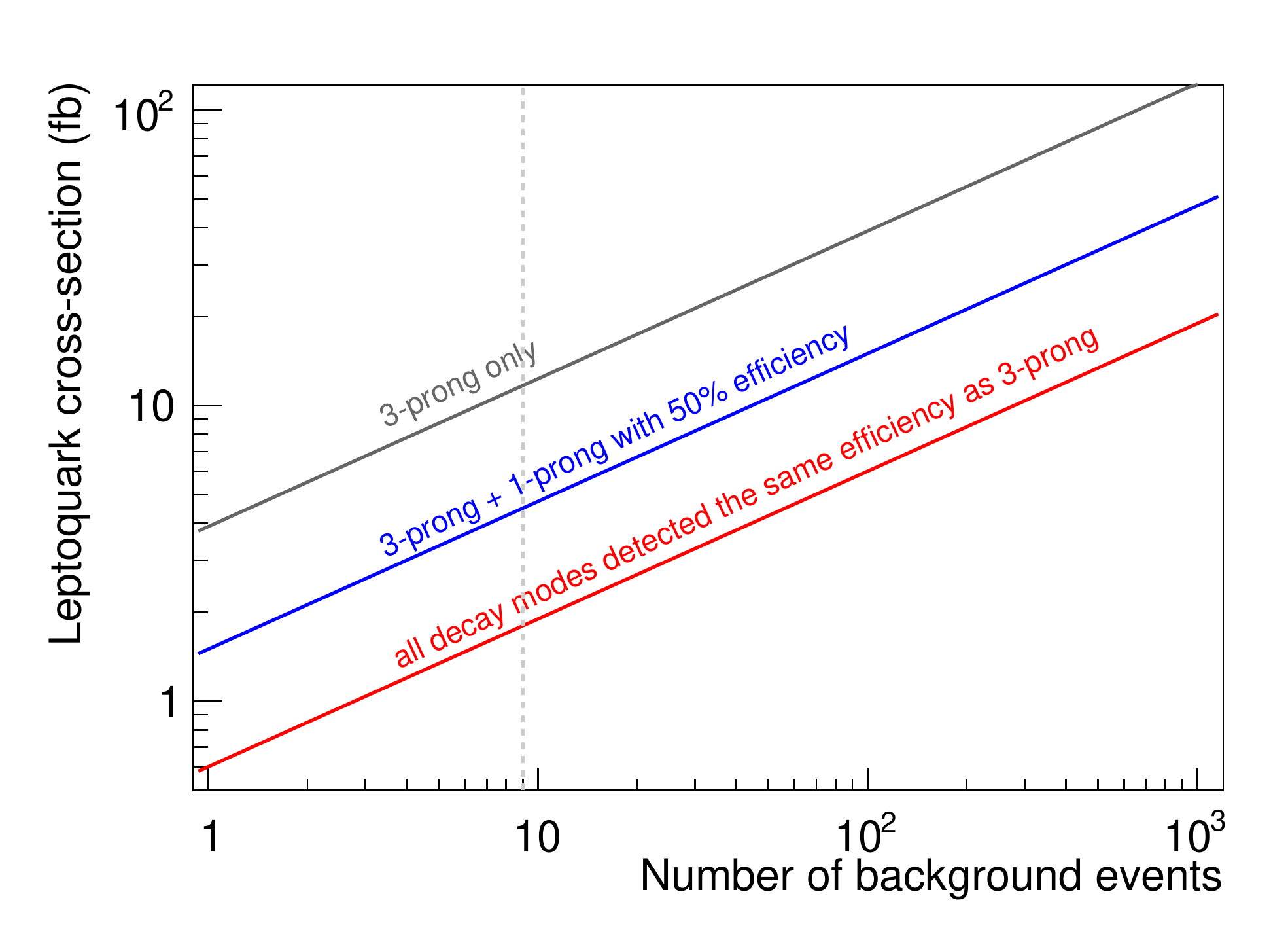} 
    \caption{Cross section sensitivity for leptoquark search vs number of residual background events for 100~fb$^{-1}$ integrated luminosity. The grey line corresponds to the scenario that 
    only ``3-prong" decay modes are detected. 
    The blue line corresponds to the scenario where 
    electron and pion ``1-prong" decay modes could be detected with 50\% efficiency of the ``3-prong" case.  
    And the red line shows the scenario if all decay modes were detected at the same efficiency as the ``3-prong'' case. 
    }
    \label{fig:LQ_xsec_bg}
\end{figure}

At the moment, instead of providing a specific estimate of the background, 
we show in Fig.~\ref{fig:LQ_xsec_bg} the leptoquark cross-section ECCE could be sensitive to as a function of the number of background events that survive the event selection for 100~fb$^{-1}$ of integrated luminosity, based on a simple calculation as follows: If the number of background events is $B$, the number of leptoquark signal events $S$ should exceed the ``3$\sigma$" limit of the expected observed background events,  $S + B > B + 3\sqrt{B}$, for the signal to be established. The corresponding cross section is scaled from the value  where only one leptoquark signal event can be observed ($S=1$).

\begin{figure}[!h]
    \includegraphics[width=0.48\textwidth]{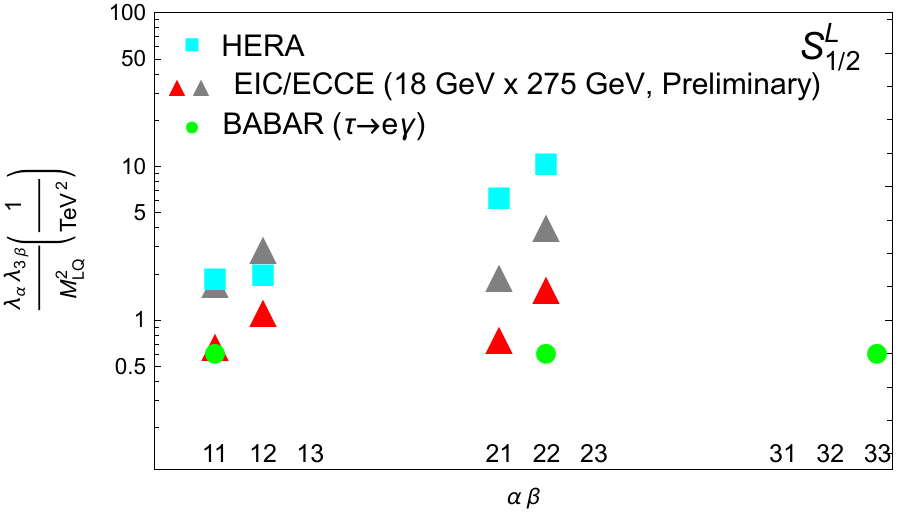}
    \includegraphics[width=0.48\textwidth]{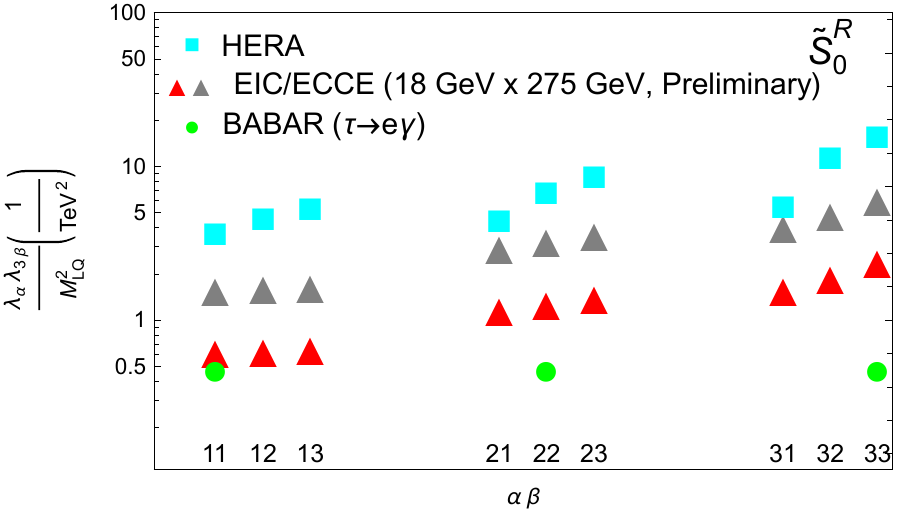}
    \caption{Limits on the scalar leptoquarks with $F=0$ $S_{1/2}^L$ (top) and $|F|=2$ $\tilde{S}_0^R$ (bottom) from 100~fb$^{-1}$ of $ep$ $18\times 275$~GeV data, based on a sensitivity to leptoquark-mediated $ep\to\tau X$ cross section of size 1.7~fb (red triangles) or 11.4~fb (grey triangles) with ECCE. Note that due to small value of $\sqrt{s}$, EIC cannot constraint the third generation couplings of $S_{1/2}^L$ to top quarks. Limits from HERA~\cite{ZEUS:2002dnh,ZEUS:2005nsy,H1:1999dil,H1:2007dum} are shown as cyan solid squares, and limits from $\tau\to e\gamma$ decays~\cite{Gonderinger:2010yn} are shown as green solid circles.}
    \label{fig:LQ_Limits_Scalar}
\end{figure}

\begin{figure}[!h]
    \includegraphics[width=0.48\textwidth]{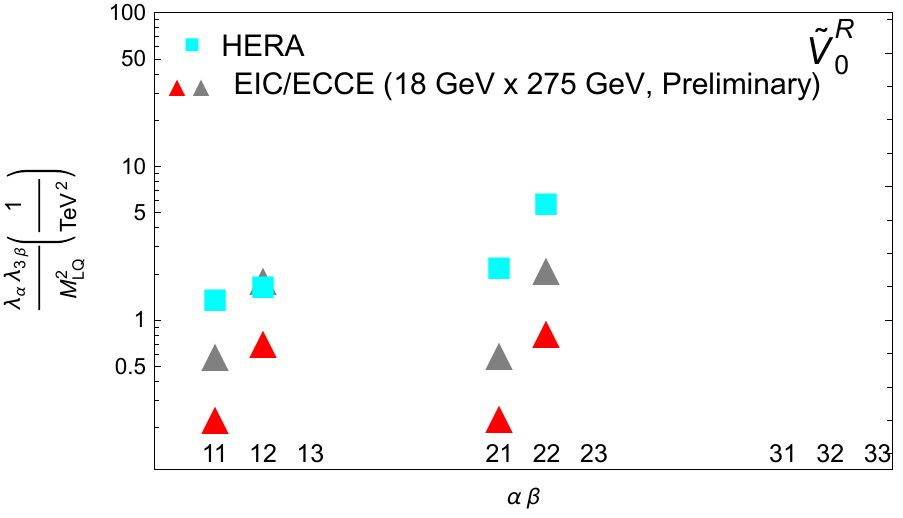}
    \includegraphics[width=0.48\textwidth]{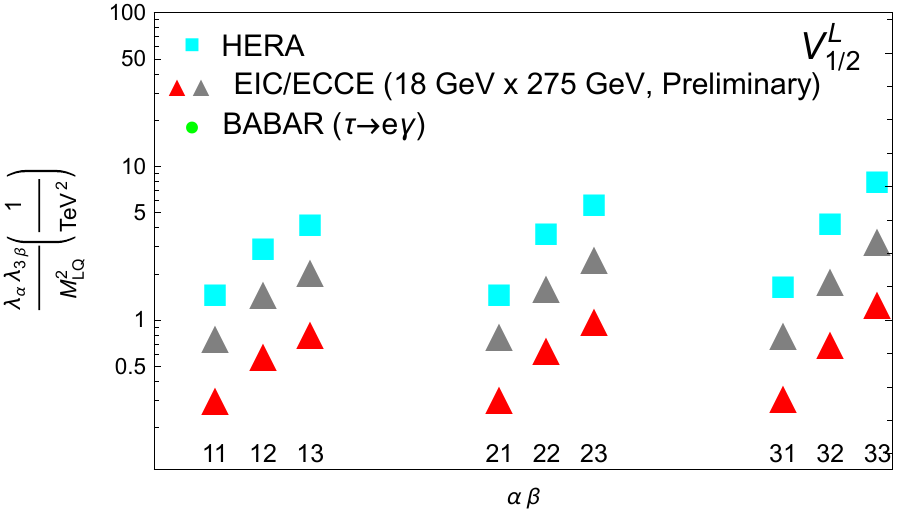}
    \caption{Limits on the vector leptoquarks with $F=0$ $\tilde{V}_0^R$ (top) and $|F|=2$ $V_{1/2}^L$ (bottom) from 100~fb$^{-1}$ of $ep$ $18\times 275$~GeV data, based on a sensitivity to leptoquark-mediated $ep\to\tau X$ cross section of size 1.7~fb (red triangles) or 11.4~fb (grey triangles) from ECCE. Note that due to small value of $\sqrt{s}$, EIC cannot constraint the third generation couplings of $\tilde{V}_0^R$ to top quarks. Limits from HERA~\cite{ZEUS:2002dnh,ZEUS:2005nsy,H1:1999dil,H1:2007dum} are shown as cyan solid squares. Limits from $\tau\to e\gamma$ decays~\cite{Gonderinger:2010yn} exist but require some work to convert to the 4-fermion contact term. This will be done in the future.}
    \label{fig:LQ_Limits_Vector}
\end{figure}

As a best-case scenario estimate of the sensitivity to the leptoquark signal cross section, we do not consider any NC and photoproduction background event since none of these events passed all the selection cuts on our limited MC event sample. For a $5\sigma$ (99.99994\% confidence level) discovery criteria of $S/\sqrt(B) \geq 5$ ($S$ being signal and $B$ being background) and use $B=9$ events from CC background, we need $S=15$ leptoquark events or a total of $15+9=24$ events to claim $e\to\tau$ CLFV discovery.  Alternatively, detection of less than $9+9=18$ events will provide a $3\sigma$ (99.7\% C.L.) exclusion limit on the leptoquark cross section, which would be $1.3$~fb$\times 3\sqrt{9}=11.4$~fb, $0.54$~fb$\times 3\sqrt{9}=5.0$~fb, and $0.19$~fb$\times 3\sqrt{9}=1.7$~fb, for detection possibility of ``3-prong only'', ``3-prong + 1-prong with 50\% efficiency", and ``all decay modes detected with same efficiency as 3-prong'', respectively. The exclusion potential, expressed in terms of $\lambda_{1\alpha}\lambda_{3\beta}/M_{LQ}^2$, are shown in Figs.~\ref{fig:LQ_Limits_Scalar} and~\ref{fig:LQ_Limits_Vector} for scalar and vector leptoquark states, respectively. This is a preliminary estimate, and different statistical methods and a larger MC event sample to better estimate NC DIS and photoproduction backgrounds could give rise to different estimates. 

  \section{Summary}
  We carried out the first projection analysis for charged lepton flavor violation in the $e\to\tau$ transition channel, using EIC simulations with the ECCE detector configuration. More work needs to be done in the future alongside the development of ECCE into a project detector, such as using detector-based particle identification, study more $\tau$ decay modes, and carry out the background study with higher statisitics.  Our current study, using the simulation and detector resources at hand, shows that the EIC will place a more stringent limit on $e\to\tau$ CLFV mediated by leptoquarks than the previous HERA data. The very high vertex resolution of the ECCE detector configuration plays a critical role in our study.

\section{Acknowledgements}
\label{acknowledgements}

We thank the EIC Silicon Consortium for cost estimate methodologies concerning silicon tracking systems, technical discussions, and comments.  We acknowledge the important prior work of projects eRD16, eRD18, and eRD25 concerning research and development of MAPS silicon tracking technologies.

We thank the EIC LGAD Consortium for technical discussions and acknowledge the prior work of project eRD112.

We thank Guillelmo Gomez Ceballos Retuerto and Hubert Spiesberger for their useful discussions and comments.

We acknowledge support from the Office of Nuclear Physics in the Office of Science in the Department of Energy, the National Science Foundation, and the Los Alamos National Laboratory Laboratory Directed Research and Development (LDRD) 20200022DR.

\bibliographystyle{elsarticle-num} 
\bibliography{EIC_EW_LQ}

\appendix
\begin{onecolumn}
  \section{Simulation Input Files and Other Details}\label{sec:appc}
 
 \subsection{LQGENEP Input Files for Leptoquark Signal Simulation} \label{sec:app_LQ_inputs}
 The input file for Leptoquark event generation using LQGENEP~1.0, for $Q^2_\mathrm{min}=10$~GeV$^2$, is shown in Fig.~\ref{fig:input_cards_lqgenep}. 
 \begin{figure}[!h]
 \centering
 \includegraphics[width=0.75\textwidth]{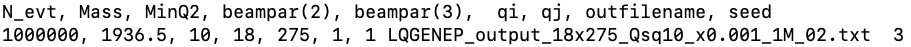}
 \caption{LQGENEP~1.0 input file for $ep$ $18\times275$~GeV, $Q^2>10$~GeV$^2$ setting.}
 \label{fig:input_cards_lqgenep}
 \end{figure}
 
  \subsection{Djangoh Input Files for DIS NC and CC Background Event  Simulation}\label{sec:app_LQ_bg_Djangoh}
  
  DIS background NC and CC events were generated with Djangoh.4.6.10 , with the input files shown in Fig.~\ref{fig:input_cards_djangoh_LQ_bg}.
  
  \begin{figure}[!h]
      \includegraphics[width=0.45\textwidth]{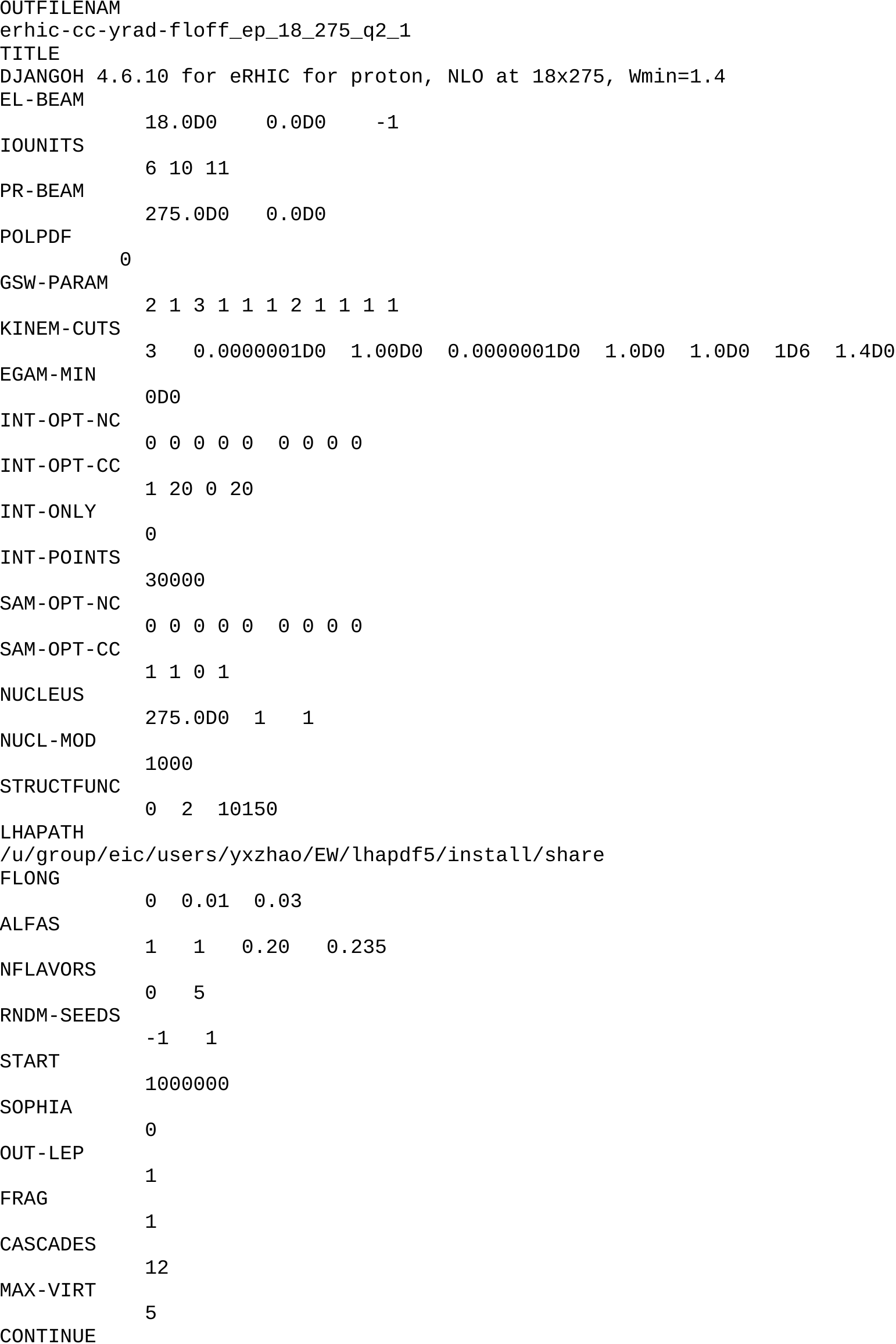}
      \includegraphics[width=0.45\textwidth]{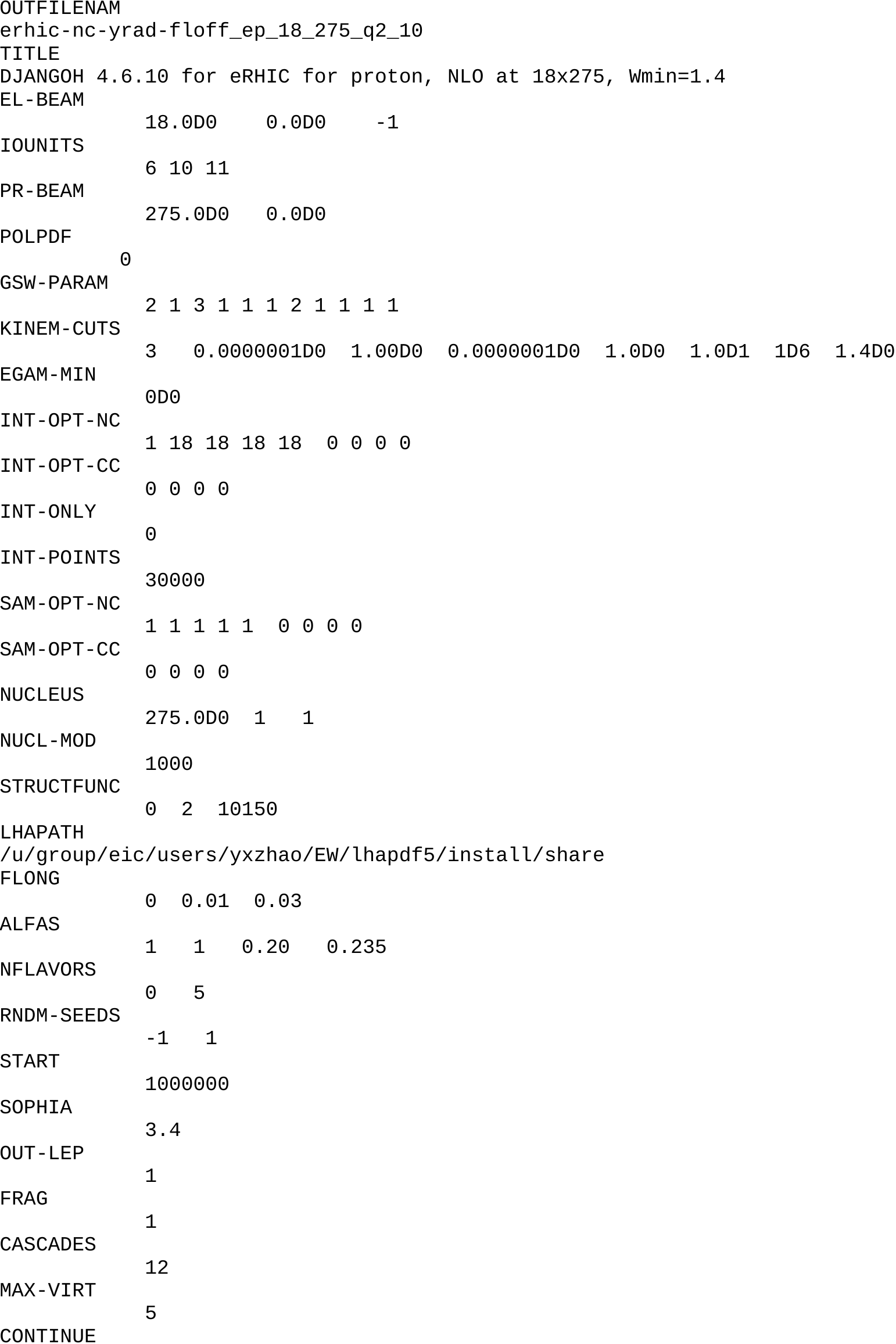}
      \caption{Djangoh (4.6.10) input files for $ep$ $18\times 275$~GeV, DIS CC (left) and NC (right) background simulations for the leptoquark study. The $Q^2_\mathrm{min}$ is set at 1~GeV$^2$ and 10~GeV$^2$ for DIS CC and NC, respectively.  }
      \label{fig:input_cards_djangoh_LQ_bg}
  \end{figure}
  
  \newpage
\subsection{Pythia Input File for Photoproduction Background Simulation}\label{sec:app_LQ_bg_pythia}
  The input file for photoproduction background generation using Pythia is shown in Fig.~\ref{fig:input_cards_pythia_photoprod}. 
  
  \begin{figure}[!h]
      \includegraphics[width=0.5\textwidth]{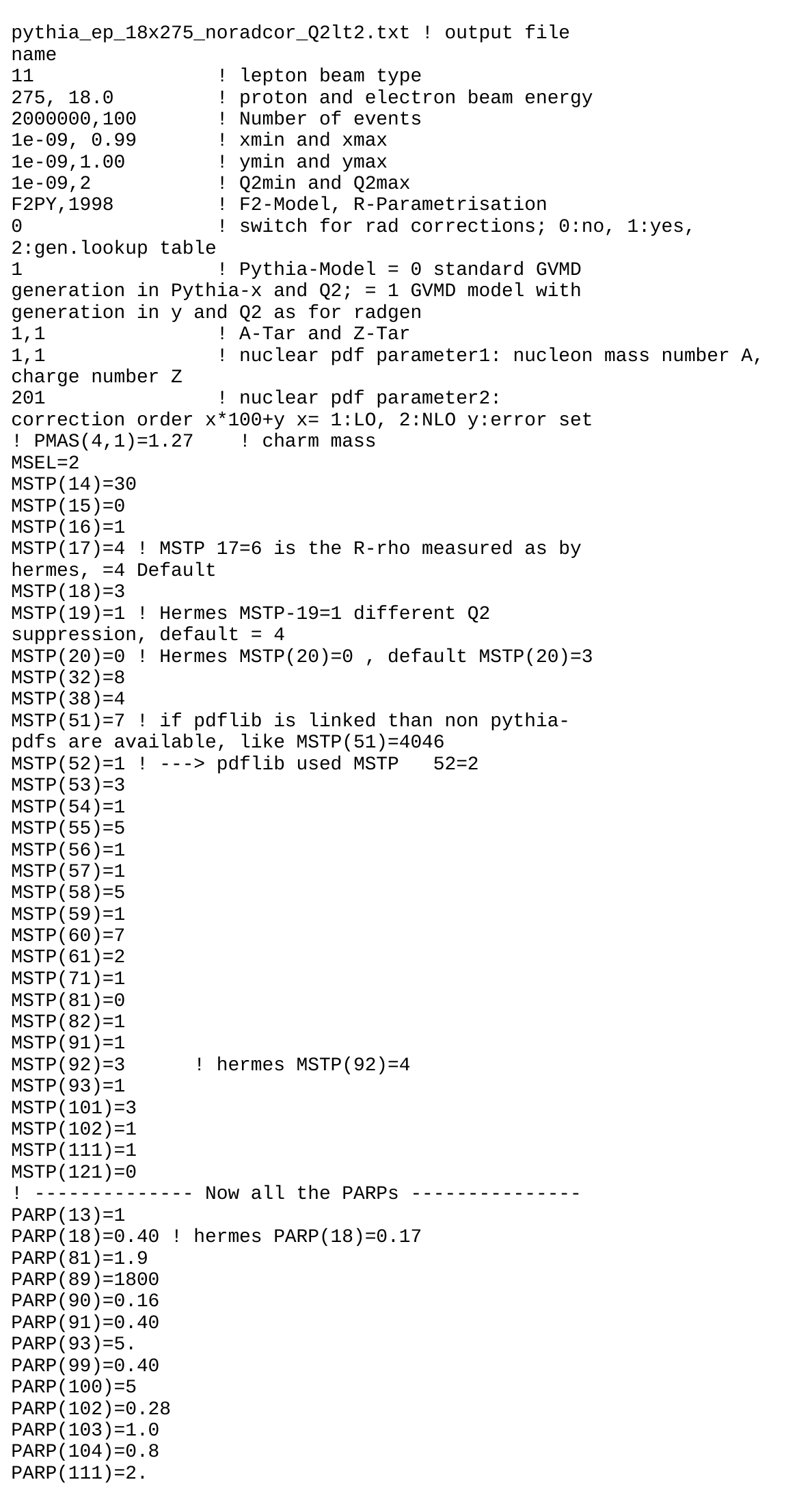}
      \includegraphics[width=0.47\textwidth]{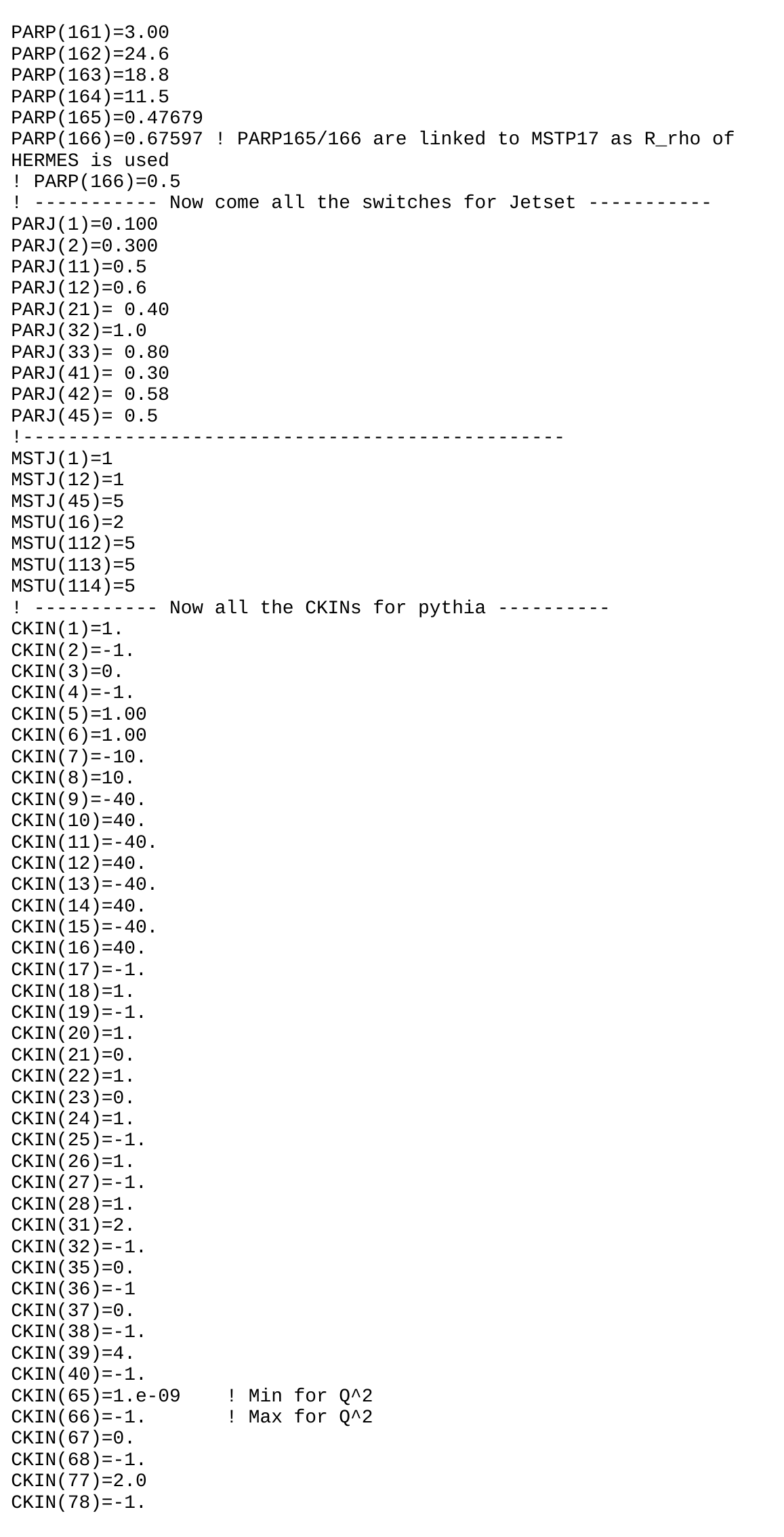}
      \caption{Pythia (6.428) input files for photoproduction for $ep$ $18\times 275$~GeV setting.}
      \label{fig:input_cards_pythia_photoprod}
  \end{figure}
  
  \end{onecolumn}

\end{document}